\documentstyle[12pt]{article}

\begin{document}

\title{Gravitation as a Super SL(2,C) Gauge Theory}

\author{Roh S. Tung\\
 {\it California Institute for Physics and Astrophysics}\\
 {\it 366 Cambridge Avenue, Palo Alto, California 94306, USA}\\
 {\it E-mail: tung@calphysics.org}}
\maketitle

\begin{abstract}
We present a gauge theory of the super SL(2,C) group. The gauge
potential is a connection of the Super SL(2,C) group. A
MacDowell-Mansouri type of action is proposed where the action is
quadratic in the Super SL(2,C) curvature and depends purely on
gauge connection. By breaking the symmetry of the Super SL(2,C)
topological gauge theory to SL(2,C), a metric is naturally
defined.
\\
\\
Proceedings of the 9th Marcel Grossmann Meeting, World Scientific.
 \end{abstract}

\newpage

Let us start with a Super SL(2,C) algebra\footnote{The upper-case
Latin letters $A,B,...=0,1$ denote two component spinor indices,
which are raised and lowered with the constant symplectic spinors
$\epsilon_{AB}=-\epsilon_{BA}$ together with its inverse and their
conjugates according to the conventions
$\epsilon_{01}=\epsilon^{01}=+1$,
$\lambda^A:=\epsilon^{AB}\lambda_B$, $\mu_B:=\mu^A\epsilon_{AB}$.
Lowercase Latin letters $p,q,...$ denote the Super SL(2,C) group
indices, $a,b,c,...=0,1,2,3$ denote the SO(3,1) indices
\cite{PR}.}
 (with three complex
SL(2,C) generators $M_{00},M_{01}=M_{10},M_{11}$ and two complex
supersymmetric generators $Q_0,Q_1$)\footnote{We can realize this
Super SL(2,C) algebra by a complex superspace $C^{2|1}$ with
coordinates $(\pi_0,\pi_1,\theta)$ where transformations are given
by $M_{AB}=\pi_A {\partial\over\partial \pi^B} +\pi_B
{\partial\over\partial \pi^A}$ and $Q_A=\pi_A
{\partial\over\partial\theta} +\theta
{\partial\over\partial\pi^A}$ with $\theta=\left(
\begin{array}{cc}
  0 & 1 \\
  0 & 0
\end{array}\right)$
 and ${\partial\over\partial\theta}=\left(\begin{array}{cc}
   0 & 0 \\
   1 & 0 \
 \end{array}\right)$.}:
\begin{eqnarray}
&&\left[ M_{AB}, M_{CD} \right] =
      \epsilon_{C(A} M_{B)D}
     +\epsilon_{D(A} M_{B)C} , \label{1}\\
&&\left[ M_{AB}, Q_C \right] = \epsilon_{C(A} Q_{B)} , \qquad
\{Q_A, Q_B\} = 2 M_{AB} , \label{2}
\end{eqnarray}
where $\epsilon_{C(A} M_{B)D}
=\textstyle{1\over2} (\epsilon_{CA} M_{BD}
+\epsilon_{CB} M_{AD})$ and 
$\epsilon_{C(A} Q_{B)} =\textstyle{1\over2} (\epsilon_{CA} Q_{B}
+\epsilon_{CB} Q_{A})$. The Super SL(2,C) group is isomorphic to
the complex extension of OSp(1,2). It is a simple super Lie group
and has a nondegenerate Killing form \cite{DeWitt}. The
Cartan-Killing metric is
$\eta_{pq}={\rm{diag}}(\textstyle{1\over2}
(\epsilon_{AM}\epsilon_{BN}+\epsilon_{AN}\epsilon_{BM}),-2
\epsilon_{AB})$.

To {\it gauge} this Super SL(2,C) group\cite{Tung2000}, we
associate to each generator $T_p=\{M_{AB}, Q_A\}$ a 1-form field
$A^p=\{\omega^{AB},\varphi^{A}\}$, and form a super Lie algebra
valued connection 1-form,
\begin{equation}
A=A^p T_p=\omega^{AB} M_{AB} +\varphi^A Q_A , \label{3}
\end{equation}
where $\omega^{AB}$ is the SL(2,C) connection 1-form and
$\varphi^A$ is an anti-commuting spinor valued 1-form.
(We shall use ${\cal D}$ for the Super SL(2,C) covariant derivative
and $D$ for the SL(2,C) covariant derivative.)

The curvature is given by $F=dA + \textstyle{1\over2}[A, A]$.
Given the Super SL(2,C) connection $A$ defined in equation
(\ref{3}), the curvature ($F=F(M)^{AB} M_{AB}+F(Q)^A Q_A$)
contains a bosonic part associated with $M_{AB}$,
\begin{equation}
F(M)^{AB}=d\omega^{AB}+\omega^{AC}\wedge \omega_C{}^B
+ \varphi^A\wedge\varphi^B ;
\end{equation}
and a fermionic part associated with $Q_A$ ,
\begin{equation}
F(Q)^A=d\varphi^A+\omega^{AB}\wedge \varphi_B .
\end{equation}

A simple action, quadratic in the curvature, using this Super
SL(2,C) connection $A$ is
\begin{eqnarray} \label{6}
 {\cal S}_{\rm{T}}[A^p]=\int\, F^p\wedge F^q \,\eta_{pq}
 =\int\,
 F(M)^{AB}\wedge F(M)_{AB} + 2 F(Q)^{A}\wedge F(Q)_{A} ,
\end{eqnarray}
where $\eta_{pq}$ is the Cartan-Killing metric of the Super
SL(2,C) group, ${\cal D}\eta_{pq}=0$. However, this action is a
total differential. Hence, similar to the work of MacDowell and
Mansouri \cite{MM}, we need to choose another spinor action which
is SL(2,C) invariant, thus breaking the topological field theory
of the Super SL(2,C) symmetry into an SL(2,C) symmetry. Let us
choose $i_{pq}={\rm{diag}}(\textstyle{1\over2} \left(
\epsilon_{AM}\epsilon_{BN} +\epsilon_{AN}\epsilon_{BM} \right),0)$
The new action (related to the quadratic spinor action
\cite{BM,NT,TJ,R1,R2,J}) is
\begin{eqnarray} \label{7}
{\cal S}[A^p]&=&\int\, F^p\wedge F^q \, i_{pq} =\int\, F(M)^{AB}
\wedge F(M)_{AB} .
\end{eqnarray}

The field equations are obtained by varying the Lagrangian with
respect to gauge potentials (the Super SL(2,C) connection). With
these gauge potentials fixed at the boundary, the field equations
are
\begin{eqnarray}
&& R^{AB}\wedge\varphi_B=0 \qquad (DF(Q)^A=0),\label{fe1} \\
&&D(R^{AB}+ \varphi^A \wedge \varphi^B)=0 \qquad (DF(Q)^{AB}=0),\label{fe2}
\end{eqnarray}
where, becuase of the SL(2,C) Bianchi identidy
($DR^{AB}=0$), the second field equation (\ref{fe2}) is reduced to
$D(\varphi^A \wedge \varphi^B)=0$.

In order to make a connection between the internal space of the
Super SL(2,C) group with the structures on the four-manifold, we
break the symmetry\cite{Kerrick,Nieto} from a Super SL(2,C)
topological field theory ${\cal S}_{\rm T}[A^p]$ into an SL(2,C)
invariant ${\cal S}[A^p]$. Using the fact that ${\cal
D}i_{pq}=C^m{}_{pn} A^n i_{mq}$ , and ${\cal D}\eta_{pq}=0$, the
metric ${\cal G}$ is defined by
\begin{equation}
{\cal G}=\eta^{pm}\eta^{qn} {\cal D}i_{pq} \otimes {\cal D}i_{mn}
=\epsilon_{AB} \varphi^A \otimes \varphi^B.
\label{spinormetric}
\end{equation}
Thus upon breaking the supersymmetry, the metric is naturally
defined.

The real spacetime metric can be obtained by considering the
complex conjugate of the generators $M_{A'B'}, Q_{A'}$ (which
satisfies the complex conjugate of (\ref{1}) and (\ref{2})), the
gauge potentials (\ref{3}) and the actions (\ref{6}), (\ref{7}).
Consequently the spacetime metric is the real part of the complex
metric ${\cal G}$, and the real tetrad $\theta^{AA'}$ is given by
$\varphi^A=\theta^{AA'}Q_{A'}$.

\smallskip
We would like to thank F. Mansouri and L. Mason for helpful
discussions.

\end{document}